\documentstyle[11pt,aas2pp4]{article}
\begin{document}
\title{Photometric Signatures of Starbursts in Interacting 
Galaxies\\
and the Butcher-Oemler Effect}

\author{Karl D. Rakos and Thomas I. Maindl}
\affil{Inst. for Astronomy, Univ. of Vienna}

\author{James M. Schombert\altaffilmark{1}}
\affil{Infrared Processing and Analysis Center\\
Jet Propulsion Laboratory\\
California Institute of Technology}
\altaffiltext{1}{Present Address: Astrophysics Division, NASA 
Headquarters, Washington, D.C. 20546}

\begin{abstract}
This paper presents new and synthetic narrow band photometry of
ellipticals, spirals, Seyferts and interacting galaxies in an attempt to
identify the cause of the unusually high fraction of blue cluster galaxies
in distant clusters (Butcher-Oemler effect).  The properties and
distribution of the low redshift sample specifically points to starbursts
as the origin of the blue narrow band colors in interacting Arp galaxies.
Comparison between theoretical models and multicolor diagrams, particularly
4000{\AA} break colors, indicates a photometric signature which differs
from both normal disk galaxy star formation and nonthermal components.
This photometric signature is absent for the Butcher-Oemler galaxies whose
general color distribution, compared to present-day clusters, is
consistent with a majority of the blue population involved in normal star
formation rates (spiral- like) with the addition of a small fraction of
bright, blue interacting/merger systems.  This photometric picture of the
Butcher- Oemler galaxies is in agreement with the morophological evidence
from HST imaging.
\end{abstract}

\keywords{galaxies: photometry - galaxies: starburst - galaxies: evolution}

\section{INTRODUCTION}

Information gleaned from spectroenergy distributions (SEDs) of galaxies are
key to our understanding of their age, metallicity, stellar population
content and star formation histories.  Analysis of SED data has usually
invoked either population synthesis, which uses a library of spectral
templates (Pickles 1985), or evolutionary models, a method of convolving
stellar isochrones with star formation rates and initial mass functions
(Bruzual 1983, Arimoto and Yoshii 1986, Guiderdoni and Rocca- Volmerange
1987).  In this series of papers (Fiala, Rakos and Stockton 1986; Rakos,
Fiala and Schombert 1988; Rakos, Schombert and Kreidl 1991; Rakos and
Schombert 1995), a system of narrow band filters has been used to replace
the SED sampling at the sacrifice of resolution for a gain in
signal-to-noise.  And, just as stellar models are tested by predictions of
integrated quantities such as total luminosity, mass, radius and effective
temperature, population synthesis results can then be tested by comparison
to predictions of integrated stellar content and metallicity of a galaxy
as given by narrow band indices.  Thus, theoretical models, which produce
spectral energy distributions as a function of age, can be convolved to
produce synthetic colors for comparison to narrow band observations at
various redshifts.

Our project has approached this problem through the use of a modified
Str\"omgren filter system called $uz,vz,bz,yz$ to distinguish it from the
original $uvby$ system.  Our primary goal was to investigate the color
evolution of ellipticals, testing the hypothesis that ellipticals are the
result of a single burst where all the stars formed at an epoch near the
time of galaxy formation and later aged through passive stellar
evolution.  However, in the previous paper in this series (Rakos and
Schombert 1995), the data revealed an extended Butcher-Oemler effect (the
increasing fraction of blue to red galaxies in distant clusters) out to
redshifts of 0.9 and containing up to 80\% of the cluster population as
blue galaxies (versus the previous values of 40\% at redshift of 0.4).
Deep HST images of intermediate redshift clusters (Dressler et al. 1994a,b)
indicate that, while the brightest blue galaxies have disturbed
morphologies suggestive of an origin of their blue colors from tidally
induced star formation, a majority of the Butcher-Oemler galaxies have
normal, late-type morphology.  The goal of this paper is to combine the
information obtained from narrow band photometry of nearby spirals,
starburst, Seyferts and interacting galaxies in order to search for
signatures of starburst activity that can then be linked to the
$uz,vz,bz,yz$ photometry of Butcher-Oemler galaxies in distant clusters.

The procedure used herein is fourfold. First, the $uz,vz,bz,yz$ system
will be linked to the original Str\"omgren system for stars to establish a
baseline stellar sequence.  Second, synthetic colors for a range of galaxy
types will be produced to map the behavior of the $uz,vz,bz,yz$ system on
galaxies with composite stellar populations and varying star formation
histories.  Third, photometry of interacting/merging galaxies will be
presented to understand the signature of starburst phenomenon in
galaxies.  And lastly, comparison of the above data will be made to high
redshift clusters to examine the hypothesis that blue cluster galaxies are
the result of similar starburst activity.

\section{PHOTOMETRIC PROPERTIES OF THE $uz,vz,,bz,yz$ COLOR 
SYSTEM}

The Str\"omgren $uvby$ filter system and its extensions (Str\"omgren 1966,
Wood 1969) were developed to provide an accurate means of measuring the
temperature, chemical composition and surface gravity of stars, without
resorting to high resolution spectroscopy, by selecting regions of the
stellar SED for narrow band photometry which are particularly sensitive to
these specific properties.  This provided an efficient and quantifiable
method of sampling stellar types without the use of qualitative spectral
classification (e.g. the MK scheme).

The $uvby$ system, modified for this project and designated as the
$uz,vz,bz,yz$ system, covers three regions in the near-UV and blue portion
of the spectrum (see Figure 1 of Rakos and Schombert 1995).  The first
region is longward of 4600{\AA}, where the influence of absorption lines
is minor.  This is characteristic of the $bz$ (4650{\AA}) and $yz$
(5500{\AA}) filters which produce a continuum color index, $bz-yz$, free
of metal and surface gravity effects.  The second region is a band
shortward of 4600{\AA}, but above the Balmer discontinuity.  This region,
covered by our $vz$ (4100{\AA}) filter, is strongly influenced by metal
absorption lines (i.e. Fe, CN) particularly for spectral classes F to M
which dominate the contribution of light in galaxies.  The third region is
a band shortward of 4000{\AA}, below the effective limit of crowding of
the Balmer absorption lines, the so-called 4000{\AA} break.  The 4000{\AA}
break has been used by several studies as a measure of the evolution of
galaxies (Hamilton 1985, Spinrad 1986, Dressler and Shectman 1987, Kimble
et al. 1989) and the $uz$ (3500{\AA}) filter is a good measure of the blue
side of this feature.  The $vz$ filter is in a good position to sample the
red side of the 4000{\AA} break with main transmission between 4050{\AA}
and 4250{\AA} and, thus, the $uz-vz$ color is a convenient measure of this
feature.

Our slight redefinition of the Str\"omgren system was primarily made to
allow for a new definition of the blue side of the $uz$ filter and to link
the photometry system with the more reliable spectrophotometry standards
that are now rapidly appearing in the literature.  The original $u$
passband in the Str\"omgren system is produced by the combination of the
ultraviolet cutoff of the Earth's atmosphere and a red blocking filter.
But, as soon as the observations concern redshifted filters (such as
distant clusters of galaxies, see Rakos and Schombert 1995), the
transmission center of the filter is moved longward and, since the
atmospheric cutoff does not redshift, the filter widens.  Our solution is
to substitute the original $u$ filter by a 200{\AA} wide interference
filter, centered at the original $u$ filter transmission peak.  However,
the interference filter is slightly narrower than the original Str\"omgren
$u$ filter and produces a slight color term with respect to the new $uz$
system (see below).

In this section, the $uz,vz,bz,yz$ system will be related to the original
Str\"omgren system for comparison to model atmospheres in the literature
and to generate an understanding of the properties of the photometry
system as it applies to the stellar populations in galaxies.  There are
two methods which can be used to derive the necessary transformations.
First, the model atmospheres of Kurucz (1979) and Bell and Gustaffson
(1978) can be convolved with the $uz,vz,bz,yz$ filters and compared to the
theoretical $uvby$ indices of those models in Lester et al. (1986).  This
first method was described in Rakos et al. (1991), but with only limited
success due to the lack of range in stellar types.

The second method is to use spectrophotometric standards where the
Str\"omgren indices are derived from spectrum scans.  This second, and
more realistic, method was performed by collecting 31 stars from the
literature with known spectrophotometry and measured Str\"omgren
photometry to form the transformations.  The sample is from the IRS
Standard Star Manual (Barnes and Hayes 1984) plus additional standards
from Gunn and Stryker (1983).  The Str\"omgren color indices are from
Hauck and Mermilliod (1980).  The resulting transformations are:

\begin{equation}
(bz-yz)=-0.268 + 0.937(b-y)
\end{equation}
\begin{equation}
mz=-0.294 + 1.092 m_1  - 0.017(b-y)
\end{equation}
\begin{equation}
cz= 0.234 + 1.034 c_1  - 0.152(b-y).
\end{equation}

\noindent Note that $m_1$ and $c_1$ are the old Str\"omgren metal-line and
surface gravity indices given as $m_1=(v-b)-(b-y)$ and $c_1=(u-v)-(v-
b)$.  The accuracy of the $mz$ and $cz$ transformations is $\pm$0.02
magnitudes.

As mentioned above, the amplitude of the 4000{\AA} break is related to the
$uz-vz$ color index.  A comparison of the break values $D_{4000}$
(Dressler and Shectman 1987) for the galaxies in \S 3 and the $uz-vz$
index reveals a linear regression of

\begin{equation}
2.5\, {\rm log}\, D_{4000} = 0.646 (uz-vz) + 0.245
\end{equation}

\noindent with a correlation coefficient of 0.91.  The 4000{\AA} break is
correlated with effective temperature, surface gravity and metallicity in
stars.  However, for an integrated stellar population, it is primarily a
temperature indicator (see Hamilton 1985 for a full discussion of the
properties of the 4000{\AA} break in stars and galaxies).  Thus, the
amplitude of the break is particularly sensitive to galaxy evolutionary
effects such as the mean age of the stellar population (position of the
main sequence turnoff point) and any recent epochs of massive star
formation (presence of hot, young stars).  This will also be the role of
the $uz-vz$ index.

\section{TWO-COLOR DIAGRAMS FOR THE $uz,vz,bz,yz$ SYSTEM}

The three primary stellar atmospheric parameters are surface gravity,
metal content and temperature.  These three parameters also dominate the
integrated light from galaxies except that surface gravity becomes the
ratio of dwarf to giant stars in the composite population and temperature
becomes the mean age of the stellar population as given by the position of
the turnoff point and the contribution of massive, young stars.  Their
combination determines our interpretation of multicolor photometry of
galaxies as guided by models of galaxy evolution (see Tinsley 1980 for a
review of the topic).  The visual light from most normal galaxies is
dominated by giants, with the low mass main sequence stars playing a minor
role (Rose 1985).  The metal content of most giant galaxies is between
solar and 0.1 of solar (Schombert et al. 1993, Burstein 1985).  In previous
papers, it has been shown that the color index $bz-yz$ is relatively free
of metal and surface gravity effects.  The $vz-yz$ index is also free of
surface gravity effects, but is sensitive to the metal content of a
galaxy.  In addition, the $uz-vz$, $uz-yz$ and $mz$ indices can be used as
a signature of hot stars (i.e. recent star formation and mean age) due to
their sensitivity to the total near-UV flux of a galaxy.

In this section, the spectrophotometric data published by Gunn and Stryker
(1983) has been used to build two-color diagrams for stars in our modified
Str\"omgren system.  Figures 1 and 2 display the synthetic colors for 162
standards of the Gunn-Stryker catalog.  For $bz-yz$ vs $vz-yz$, an
excellent correlation exists between these two colors, as would be
expected, but this correlation is surprisingly independent of the mean
surface gravity (i.e. stellar type).  Note that the reddening vector is
parallel to the star sequence.  The derivation of the reddening vector
according to Sinnerstad (1980) and Crawford (1975) is

\begin{equation}
E(c_1) = 0.20 E(b-y)
\end{equation}

\noindent and

\begin{equation}
E(m_1) = -0.32 E(b-y)
\end{equation}

\noindent from which it can be shown that

\begin{equation}
E(mz) = -0.39 E(bz-yz)
\end{equation}
\begin{equation}
E(cz) = 0.06 E(bz-yz)
\end{equation}
\begin{equation}
E(uz-vz) = 0.67 E(bz-yz)
\end{equation}
\begin{equation}
E(vz-yz) = 1.61 E(bz-yz)
\end{equation}
\begin{equation}
E(bz-yz) = 0.72 E(B-V).
\end{equation}

\noindent The ratio of total to selective absorption is 3.3, according to
Moreno and Moreno (1975), and it follows that

\begin{equation}
A(V) = 4.58 E(bz-yz).
\end{equation}

\noindent Thus, the reddening vector is parallel to the stellar sequence
for $vz-yz$ vs $bz-yz$ and slightly orthogonal for $uz-vz$ vs $bz-yz$.

For these color indices, metallicity effects are more dominant than
surface gravity effects.  Unfortunately, spectrophotometry of stars with
low metal content is rare.  To investigate the variance with metallicity,
35 stars were collected from the literature with the $[Fe/H]$ equal or
less than $- 1.5$ (Olsen 1983, Francois 1986, Magain 1987 and Tourkin
et al. 1985) and their spectrophotometry was transformed to our color
system.  Within the range $0.2 < bz-yz < 0.8$ the color indices form a
linear sequence such that

\begin{equation}
(vz-yz) = -0.227 + 2.524 (bz-yz)
\end{equation}

\noindent with a correlation coefficient of 0.97 (see solid line in the
Figure 1).  The difference in Figure 1 between the solar metallicity stars
from the Gunn and Stryker spectrophotometric catalog is evident, but not
overwhelming.  Similar deviations for stars of different metal content in
$mz$ index is also present, but this was previously known from Str\"omgren
photometry studies (see Gustaffson and Bell 1978).  Therefore, metallicity
effects are not expected to strongly influence the discussions on star
formation histories of galaxies.

The second set of two-color diagrams is the $uz-vz$ vs $bz-yz$ diagram
also shown in Figure 1.  Low temperature red giants are well separated
from dwarfs in Figure 1, but the stellar types become blurred for colors
bluer than $bz-yz=0.4$.  The behavior of stellar types in this diagram is
similar to broadband two color diagrams (e.g. Johnson $UBV$) with its
classic ``S'' shaped sequence.  The third and fourth set of two-color
diagrams are $bz-yz$ vs $mz$ and $uz-vz$ vs $mz$, where the distributions
are identical for the various stellar types.  However, note that the
reddening vector is orthogonal to the stellar sequences in both diagrams.

For comparison to the stellar sequences, Figures 3 and 4 display the same
diagrams for bright galaxies of various Hubble types.  The galaxy data are
computed synthetic colors from high resolution, high signal-to-noise
spectra published by Gunn and Oke (1975), Yee and Oke (1978), De Bruyn and
Sargent (1978), Kennicutt (1992a) and Ashby et al. (1992).  The galaxies
are broken into four classes; ellipticals, spirals, Seyferts and
starbursts, and plotted as separate symbols in Figures 3 and 4.  In
general, for the $vz-yz$ vs $bz-yz$ plane, normal galaxies (ellipticals
and spirals) follow the stellar sequence outlined in Figure 1.  Seyfert
galaxies are bluer than spirals (the bluest objects are all Seyferts), but
lie near the stellar sequence.  Note that the spectrophotometry from which
these synthetic colors are based, have been observed with small aperture
or slit sizes.  Thus, these values are strongly influenced by the
nonthermal continuum component in Seyferts (the filters avoid most of the
strongest emission lines).  Lastly, starburst galaxies deviate from the
trend for normal and Seyfert galaxies by lying well below the stellar
sequence, having bluer $vz-yz$ colors for their $bz-yz$ index (i.e. extra
UV flux).

Dwarf stars are numerous in galaxies and are a dominate portion of the
mass, but their integrated light has a minor influence on the total color
of a typical bright galaxy (Rose 1985).  The $uz-vz$ vs $bz-yz$ diagram is
the most sensitive to changes in stellar type (main sequence to giants).
However, the galaxy data displays little indication of any change in the
dominate stellar type with respect to the integrated light.  No
differences between ellipticals and spirals are distinguishable even
though there are extensive differences in their stellar populations.

For a reddening comparison, the position of a B0I star is marked in
Figures 3 and 4 along with the reddening vector for that same star under 3
magnitudes of extinction.  The data in Figures 3 and 4 confirms results
from previous studies of starburst galaxies that their position in the
$uz-vz$ vs $bz-yz$ diagram is similar to that of an object with a very
young, hot stellar population shrouded in gas and dust.  Under this
interpretation, many of the starburst galaxies show a intrinsic reddening
of between 0.5 to 3 mags of extinction.

In both $mz$ plots for galaxies (see Figure 4), the starburst galaxies are
well separated from the normal ellipticals, spirals and Seyferts, and are
distinguished by consistently having $mz$ values of less than $-0.2$.  As
with Figure 3, the reddening vector is shown with an origin of a B0Ia star
and 3 magnitudes of extinction.  It is clear from these plots that the
$mz$ index provides the best signature for a starburst phenomenon in
galaxies and both Figures 3 and 4 suggest that the narrow band optical
properties of starburst galaxies are dominated by a strongly reddened,
young stellar population, as one would expect from their far-IR properties
(see Devereux 1989).

The reddening of individual galaxies can be estimated using reddening free
parameters introduced by Str\"omgren.  The $uz,vz,bz,yz$ photometric
system has similar expressions:

\begin{equation}
[bz-yz] =  (bz-yz) - 0.66 E(B-V)
\end{equation}
\begin{equation}
[uz-vz] = (uz-vz) - 0.67 E(bz-yz)
\end{equation}
\begin{equation}
[vz-yz] = (vz-yz) - 1.61 E(bz-yz)
\end{equation}
\begin{equation}
[mz] = mz + 0.39 E(bz-yz)
\end{equation}
\begin{equation}
[cz] = cz - 0.06 E(bz-yz).
\end{equation}

\noindent In this manner, an estimate of the unknown amount of reddening
can be made and, therefore, our indices will only be dependent on the age
and the total mass of the starburst.  These are still highly degenerate
quantities, however, the color indices can be decoupled from the
non-evolutionary properties of the galaxy.

\section{$uz,vz,bz,yz$ INDICES AND STARBURST MODELS}

The dependence on age and burst strength of the underlying stellar
population can be calibrated using theoretical models of starburst
galaxies.  To this end, the SED models of Leitherer and Heckman (1995)
have been used to construct synthetic colors in order to follow changes in
the $uz,vz,bz,yz$ color indices with the time.

To illustrate the changes in our four colors ($mz$, $uz-vz$, $vz-yz$ and
$bz-yz$) two different models were considered.  The first is a combination
of a standard elliptical ($mz=-0.02$, $uz-vz=0.89$; $bz-yz=0.35$ and $vz-
yz=0.68$, see Schombert et al. 1993) with a starburst of total mass $10^6
M_{\sun}$, solar metallicity, a Saltpeter IMF ($x=2.35$) and an upper
mass limit of 100 $M_{\sun}$.  The evolution of the colors is shown in
Figure 5 for three cases; one where the elliptical galaxy has the same
total luminosity as the starburst (solid line), the second where the
elliptical galaxy is one magnitude fainter than the starburst (dotted
line) and the third where the elliptical galaxy is one magnitude brighter
than the starburst (dashed line).  Note that the luminosity, rather than
burst mass, is used for normalization and the brightness of the burst is
taken as the peak luminosity at 5500{\AA} immediately after the start of
the burst (t = $10^6$ years).  Several obvious trends are visible in
Figure 5.  All the color indices move progressively from blue to red as
the burst ages, although this progression is not necessarily smooth due to
subtleties in the combination of the stellar tracks of various stellar
masses.  For example, there is a red bump at log $t = 7.9$ due to a
post-AGB phase.  Another notable point is that both $vz-yz$ and $bz-yz$
approach the red color of an elliptical by $10^8$ years of the initial
burst, making starburst detection difficult in a relatively short time.

Surprisingly, the 4000{\AA} break color, $uz-vz$, is the least sensitive
index for detecting a past burst event.  The $uz-vz$ color approaches an
elliptical value after only approximately $10^7$ years, independent of
burst strength relative to the luminosity of the underlying population.
Figure 1 in Rakos and Schombert (1995), where the spectrum of an old
composite population (17 Gyrs) is compared with one of a young population
(1.5 Gyrs), depicts this effect.  A strong, intermediate age population
actually produces more flux in the $vz$ band relative to the $uz$ band due
to a heavy Balmer contribution from A stars.  This, in turn, implies 
rapid changes $uz-vz$ index once the O and B stars evolve to become red
giants.  This would also explain why the 4000{\AA} break has produced
mixed results as an indicator of galaxy evolution in low to intermediate
redshift samples (see Dressler 1987) since an artificial reddening of the
index would appear to mimic normal ellipticals if other colors were not
measured for comparison.  This is also evident in our previous work on
high redshift cluster blue galaxies (Rakos and Schombert 1995).  Figure 5
of that paper demonstrates that although the UV flux of this population
changes dramatically beyond $z=0.6$ (as measured by $uz-yz$), the $uz- vz$
colors are indistinguishable from the lower redshift galaxies.  The red
rise in $uz-vz$ colors at log $t=7.3$ has a corresponding blue dip in
$vz-yz$ and $bz-yz$ and indicates that blue galaxies are better selected
by  longer baseline continuum colors rather than 4000{\AA} colors.

The $mz$ index is the most sensitive of all our indices to the existence
of an underlying  burst population having the largest magnitude change
during the aging of the burst.  And, significant differences are visible
up to $5\times10^8$ years, providing a greater range of detection for the
$mz$ index.  The sharp red bump at $10^8$ years is due to a sudden onset
of post giant branch supergiants from the initial burst, however, its
time phase is small.  Notice this bump is stronger for a stronger burst
due to a larger number of intermediate mass stars climbing the AGB.

The second experiment with the theoretical models of Leitherer and Heckman
is the combination of a spiral, with constant star formation, and a burst
as shown in Figure 6.  In this example, the SB galaxy NGC 3227 was
selected.  NGC 3227 displays a moderate level of star formation (Kennicutt
1992b) with nominal colors of $mz=-0.05$, $uz-vz=0.55$, $bz-yz=0.22$ and
$vz- yz=0.39$.  The three curves in Figure 6 are the same as in Figure 5
with the initial burst set to 1 mag below the luminosity of the spiral at
5500{\AA} (dashed line), equal in luminosity (solid line) and 1 mag above
(dotted line).  The trends in Figure 6 are the same as was found for an
elliptical plus star burst combination, however, the separation between
the various burst strengths is less distinct in the early epochs due to
the fact that the underlying population is already fairly blue from the
constant disk star formation typical of a spiral.  Again, note that the
$uz-vz$ colors redden faster than any of the other filter combinations and
that the stronger the burst, the redder the peak at log $t=7.3$, opposite
to initial expectations for blue filters.  Also, as with the elliptical
plus starburst model, the $mz$ index is the most sensitive to detecting a
past starburst event.  For both the elliptical and spiral underlying
models, $mz$ has the shallowest slope with time and will show evidence of
a past starburst long after the other color indices have approached
quiescent colors.

\section{NARROW BAND PHOTOMETRY OF ARP/MARKARIAN GALAXIES}

To confirm the previous sections results based on synthetic colors from
spectrophotometry, new narrow band photometry, in the same filter sets as
used for our high redshift work, was obtained for 13 Arp and Markarian
galaxies.  The sample, listed in Table 1, was selected based either on
their far-IR, submillimeter or emission line properties to be systems of
extremely high current rates of star formation and/or high molecular gas
content, or by having a disturbed morphology suggestive of a recent merger
of a gas-rich spiral (see Schombert et al. 1990).  Table 1 presents the
galaxies selected where column 1 displays the galaxy name, column 2
contains the galaxy coordinates (epoch 1950) and column 3 displays the
galaxy redshift.

The CCD data for the Arp/Markarian sample were obtained on the Lowell
Observatory Hall (1.8m) and Perkins (1.1m) telescopes.  The imaging device
on the 1.8m telescope was a TI 800 by 800 CCD with an image scale of 0.86
arcsecs per pixel.  The 1.1m telescope imaging device was a Tektronix 512
by 512 CCD also with an image scale of 0.86 arcsecs per pixel.  The 1.8m
observations were obtained on 1-4 Oct 1993.  The 1.1m observations were
obtained 23-28 Jun 1994.  All the Arp objects were imaged on the 1.8m
telescope, the brighter Markarian galaxies were imaged on the 1.1m
telescope.  Exposure times varied from 150 to 300 secs under photometric
conditions with multiple frames per filter being co-added to increase the
signal-to-noise and eliminate cosmic-ray events.  Twilight flats were
obtained for all four filters.  The data were calibrated for both runs
using the spectrophotometric standards HD 19445, HD 192281, Feige 15, BD
+40 4032, BD +20 4211, HD 161817 and BD +26 2606 (Massey et al. 1988).
Aperture photometry, centered on regions of peak intensity, was performed
using the IRAF reduction package after erase line subtraction and
flattening.  Corrections for Galactic reddening followed the prescription
of Burstein and Heiles (1978).  Small k-corrections to final colors follow
the procedure outlined in Fiala, Rakos and Stockton (1986).   For the
redshifts of the selected galaxies (see Table 1), these corrections were
always less than 0.02 in $mz$.  Isophotal maps from the $yz$ images are
presented in Figure 7.

In addition to the photometry of selected regions and the isophotal maps
in Figure 7, $yz$ surface brightness profiles of each galaxy were also
analyzed.  Due to the well-known irregular isophotal shapes of the sample
galaxies, ellipse fitting was not practical or meaningful.  Instead, a
procedure of extracting wedge cuts along the minor and major axis was
used.  The purpose of these profiles is a simple comparison to an
exponential or $r^{1/4}$ shape, not a full structural analysis.  As an
illustration, the profile for Arp 169 B and C, two $r^{1/4}$ profiles, is
plotted in Figure 8.

The photometry can be found in Table 2 where the galaxy region designation
from Figure 7 is listed in column 1, aperture size in pixels in column 2,
$uz-vz$ color in column 3, $vz-yz$ color in column 4, $bz-yz$ color in
column 5 and $mz$ in column 6.  Aperture photometry is used in this study
since the synthetic colors from spectrophotometry are predominately from
small slits or apertures centered on the highest surface brightness
regions.  We also used primitive circular aperture photometry for our high
redshift clusters due to the lack of spatial information, so comparison to
those samples necessitates a similar reduction method.  Typical errors
where 0.05 in $uz-vz$, 0.02 in $bz-yz$, 0.03 in $vz-yz$ and 0.04 in $mz$.
The following are remarks concerning photometry of the individual galaxies
in the sample:

{\bf Arp 81}:  Figure 7 shows the two interacting galaxies that make up
this system.  The galaxy labeled E has the photometric signature of
regular E galaxy in terms of mean colors and a surface brightness profile
that is pure $r^{1/4}$.  The second galaxy B has the appearance of a bared
spiral disturbed by a neighbor.  At the center, the galaxy has a nuclear
hotspot that the $uz$ filter indicates is composed of an unusual number of
hot stars.  The region A is also active (blue in mean color) but not
within the definition of a strong starburst phenomenon, but rather a
buried AGN is indicated (see Bushouse 1987).

{\bf Arp 112}:  Both galaxies A and B display signatures of a star forming
system and strong intrinsic absorption based on their positions in the two
color diagrams.  The galaxy A has exponential surface brightness profile.
Galaxy B has an irregular profile shape.

{\bf Arp 118}:  Both galaxies A and B have general $r^{1/4}$ profile
shapes with some irregularity present.  Colors are characteristic of a
starburst in its late stages (greater than $10^8$ years).  The central
region is known to harbor a Seyfert type 2 nuclei (Hippelein 1989).

{\bf Arp 157}:  Galaxy A appears unobscured, however, galaxy C is hidden
behind a dust lane, consistent with their colors.  Both are involved in an
intense starburst event.  The galaxy A has a partly disturbed $r^{1/4}$
profile.  The whole system is a luminous IRAS source (Young et al. 1989).

{\bf Arp 169}:  Both galaxies B and C are typical ellipticals without any
signature of recent star formation in their mean colors (see also
Schombert et al. 1990).  The brightness profiles for both systems are pure
$r^{1/4}$ (see Figure 8).

{\bf Arp 182}:  The photometry in the $uz$ band was disturbed by weather.
However, galaxies A and B show the presence of hot stars and dust in the
centers based on the redder filters.  Galaxy A has disturbed
brightness profile, B has a $r^{1/4}$ profile outside the inner 5
arcsecs.

{\bf Arp 212}:  Arp 212 is a well-studied starburst galaxy (Devereux 1989)
with exponential profile (disk-like).  The narrow band colors indicate a
strong starburst with equally strong dust extinction in the bulge as
confirmed by far-IR emission.

{\bf Arp 223}:  Although Arp 223 has a disturbed morphology (a classic
shell/merger galaxy, see Schombert et al. 1990), its narrow band colors are
similar to a normal elliptical.  Given the structural signatures of a
recent merger, Arp 223 is a probable late stage starburst with a mean age
greater than $10^8$ years.

{\bf Arp 278}:  Unfortunately, the $uz$ observations for this galaxy were
lost to weather.  However, the starburst signature is present in many of
the bright knots embedded in both galaxies based on $vz$, $bz$ and $yz$
colors.  Knot G has, in comparison to other regions in the interacting
system, the highest reddening due to dust.

{\bf Arp 282}:  This galaxy pair has dramatically different colors.
Galaxy A has very blue, star-forming colors.  Galaxy B displays the colors
of a reddened elliptical, yet has a disk morphology.

{\bf Mrk 309}:  This galaxy is known in the literature as a typical WR
galaxy (Conti 1991) and displays the bluest colors of our sample.  It is a
blue compact galaxy with a $r^{1/4}$ profile.  A majority of the surface
area of the galaxy is involved in massive star formation.

{\bf Mrk 480}:  This system is most probably a merging galaxy pair, both
disk galaxies based on their exponential profiles.  Both galaxies show
fading starburst signatures of intermediate mass B and A stars (Mazzarella
and Boroson 1993).

{\bf Mrk 496}: This galaxy has two compact, blue nuclei with only moderate
intrinsic absorption based on very blue $uz-vz$ colors.  The surface
brightness profile is exponential in the interior and irregular at the
edge.  This galaxy is an ultraluminous IRAS source (log
$L_{FIR}/L_{\sun}$ = 11.33, Condon et al. 1991).

Figures 9 and 10 display the photometry for the Arp/Mrk sample in the same
fashion as Figures 1 through 4.  The photometry for different apertures of
the same objects are connected by lines.  Also shown on the diagrams are
boxes that outline the 90\% regions for ellipticals, Seyferts and
starburst galaxies from Figures 3 and 4.  There is a small trend of blue
gradients in $vz-yz$, $bz-yz$ and $mz$.  The exception to this trend is
found in the $uz-vz$ vs $bz-yz$ diagram where redder $uz-vz$ indices are
found for bluer $bz-yz$ (see discussion in \S 4).

The main result from Figures 9 and 10 is that a majority of the Arp/Mrk
objects occupy the same region of the two color diagrams that the
starburst sample defined.  This confirms our interpretation of the narrow
band colors, particularly the $mz$ differences for starburst galaxies, in
their ability to detect a photometric signature of a starburst versus AGN
activity or normal disk star formation (constant SFR).  A significant
fraction (25\%) have elliptical-like colors and profile shapes (i.e.
$r^{1/4}$), however, it should be noted that several of the Arp objects
appear to be interactions between ellipticals and spiral disks.  In Figure
10, the elliptical plus starburst tracks from Figure 5 are plotted in the
$mz$ plane (dotted line corresponding to the equal luminosity model).
Many of the starburst colors are consistent with ages of $10^7$ years or
younger, if notable reddening is present, which is in agreement with
dynamical models of the star formation history of interacting galaxies
(Mihos and Hernquist 1994).  The $mz$ vs $uz-vz$ diagram indicates that
many of the red objects could be old starbursts with ages of greater than
$10^8$ years.  Outside of blue elliptical colors and weak starburst
indices, only one galaxy (Arp 81A) displays consistent Seyfert colors (Arp
118, a known Seyfert 2, lies on the boundary between ellipticals and
Seyferts).  The lack of Seyfert colors is probably due to contamination
and mixture of stellar populations that occurs with the wider aperture
sizes compared to the narrower slits of the spectroscopy surveys.

\section{STARBURSTS AND THE BUTCHER-OEMLER EFFECT}

From the theoretical models in \S 4 plus the synthetic colors from
spectrophotometry and new photometry of Arp/Mrk galaxies presented in \S
5, a fairly consistent picture is outlined for the photometric signature
of a starburst event, old or new, in a composite stellar population.
Although blue $vz-yz$ and $bz-yz$ colors are found in starburst galaxies,
reddening distorts the interpretation of the colors with respect to age
and strength of the burst.  The $uz-vz$ or 4000{\AA} index is a good
measure for very young starbursts, however, intermediate mass stars
produce artificially red colors (see \S 4).  The most powerful indicator,
as seen in Figure 4, is the $mz$ index where a value less than $-0.2$
indicates a starburst regardless of reddening.  The $mz$ index is also the
most sensitive to detecting old starbursts hidden within an elliptical
population since it is more sensitive to turnoff stars and evolves slowly
from the time of the initial burst (see Figure 5).

Figure 11 displays the $mz$ values for the ellipticals, spirals, Seyfert
and starburst spectrophotometry of \S 3 (mean values are indicated by an
arrow at the top of each histogram).  Although there is some overlap in
the range of $mz$ values for each galaxy type, the ellipticals have the
reddest $mz$ value of 0.02 and starbursts fall at the blue end of $-
$0.29.  Spirals and Seyferts have mean values between $-$0.05 and
$-$0.10.  Also shown in Figure 11 are the data from Rakos and Schombert
(1995) for the distant cluster sample divided into two redshift bins at
the top right, and the $z>0.6$ galaxies divided by $bz-yz$ color into red
and blue galaxies (see that paper for a full description of the sample).
The Butcher-Oemler galaxies ($z>0.6$ blue galaxies in the bottom right
panel of Figure 11) have a mean color indistinguishable from the mean for
the spiral sample.  However, the distribution of Butcher-Oemler galaxies
differs from the normal spirals distribution by having a notable blue
tail.  We also note that the bluest Butcher-Oemler galaxies are often the
brightest galaxies in their respective clusters.

The extreme blue members of this color distribution must attribute their
colors to a starburst phenomenon since neither spirals or Seyferts in the
low redshift sample display these colors.  This is an important
distinction since spectroscopy surveys of distant clusters have isolated
a high fraction of cluster galaxies with strong emission lines (Dressler
and Gunn 1983) which offers an alternative explanation of the
Butcher-Oemler effect in terms of either high levels of current star
formation (i.e. strong emission from HII regions) or increased AGN
phenomenon with higher redshift.  Our $uz,vz,bz,yz$ filter system
specifically avoids the dominant emission lines in AGN activity and
focuses on the underlying continuum emission from the hot gas and stellar
populations.  Thus, the lack of agreement in $mz$ with the Seyfert samples
would discount the AGN effects in favor of strong star formation as the
explanation for the Butcher-Oemler effect.

This photometric result is consistent with the recent morphological
results from HST imaging of $z=0.4$ clusters by Dressler et al. (1994a,b).
In that study it was found that, while some of the brightest blue galaxies
showed evidence of enhanced star formation due to tidal effects or
mergers, most of the blue cluster galaxies are normal, late-type spirals
and irregulars, their blue colors being due to normal star formation
rates and histories.  The number of disturbed systems, ones which had
clearly undergone a recent tidal encounter or full merger, is higher in
these distant clusters as compared to present-day cluster populations, but
represent less than 10\% of the cluster population.  Dressler et al.
(1994b) also notes that the morphologically disturbed systems are the
brightest blue galaxies, supporting a starburst interpretation for their
origin.

The morphological information combined with our photometric results
indicated that the color distribution of the Butcher-Oemler galaxies in
Figure 11 is due to two components.  The first is a sequence of
morphologically late-type galaxies with normal star formation rates and
histories.  The second component is a blue tail composed of galaxies with
starburst colors and disturbed morphologies.  We also note that the
$uz-vz$ or 4000{\AA} break colors of the high redshift blue cluster
galaxies are not as strongly blue as compared to the continuum colors
(e.g. $uz-yz$ or $bz-yz$).  The theoretical models of Leitherer and
Heckman (1995) demonstrate that this is the result of a increasing
contribution from intermediate mass stars (A and F type) as the starburst
ages.  This produces artificially red 4000{\AA} break colors when, in
reality, the UV flux is still quite high from young stars.  This would
also explain Hamilton's (1985) observation that there is little evolution in
the 4000{\AA} break colors of high redshift, color-selected field galaxies
in terms of a lack of high mass O and B stars rather than a lack of color
evolution.

\section{CONCLUSIONS}

Rakos and Schombert (1995) demonstrated that Butcher-Oemler effect is much
more dramatic than previously thought.  Early work on cluster galaxies found
blue fractions from 30 to 40\% in clusters with redshifts of 0.4.  Rakos
and Schombert found fractions of 80\% in clusters at redshifts of 0.9.
This is a very dominate effect on cluster populations not only in terms of
the rapid pace of galaxy evolution as first observed by Butcher and
Oemler, but also in the that extent of the blue galaxy population prevails over
the entire cluster population at high redshifts (80\% would imply all
present-day S0's are actively star-forming at these epochs).

In this paper we have derived and described the fundamental properties of
the $uz,vz,bz,yz$ filter system. In particular, we have demonstrated the
advantage of $mz$ index as a detector of starbursts in galaxies.  We then
applied this index to establish the photometric signature of starbursts in
interacting galaxies.  This method was then applied to high redshift
clusters to demonstrate that the Butcher-Oemler galaxies are due to a
mixture of star-forming spirals and irregulars plus an addition component
from a small fraction of starburst galaxies.

The following is a summary of our primary results and interpretations:

\noindent (1) The $uz,vz,bz,yz$ filter system is a robust tool for
exploring the stellar populations in galaxies due to the placement of the
filter centers in regions of clearer evolutionary interpretation.  The
narrow width of the filters avoids contaminating emission lines so as to
concentrate on the underlying continuum emission from stellar atmospheres
or nonthermal emission.

\noindent (2) The $uz,vz,bz,yz$ system is relinked to spectroscopy
standards and the transformations from the original Str\"omgren system are
derived.  The reddening effects are outlined and displayed in Figures 1
and 2.

\noindent (3) Using spectrophotometry of ellipticals, spirals, Seyferts
and starburst galaxies from the literature, synthetic $uz,vz,bz,yz$
indices are produced and plotted on multi-color diagrams.  Specific
regions are occupied by various galaxy types and the $mz$ index is
demonstrated as the best indicator of starburst activity.

\noindent (4) Theoretical models from Leitherer and Heckman are used to
define the behavior of the $uz,vz,bz,yz$ indices with time from the
initial starburst.  The inadequacy of the 4000{\AA} break color, $uz- vz$,
is shown due to the contribution of intermediate mass stars to an aging
burst.  Again, the $mz$ index is shown to be the most sensitive to
starburst activity and for the longest amount of time post-starburst.

\noindent (5) New photometry for a sample of Arp and Markarian galaxies is
presented and confirms the results from the synthetic colors that
starburst galaxies have a unique, detectable signature in the
$uz,vz,bz,yz$ system.  Selection by disturbed morphology also confirms the
dynamical arguments for starbursts in merging/interacting galaxies.

\noindent (6) Direct comparison with the photometry of blue galaxies in
distant clusters ($z>0.6$) yields the interpretation that the Butcher-
Oemler effect is due primarily to normal star formation with a small
component due to tidally induced starbursts.  The recent results from HST
imaging (Dressler et al. 1994a,b) confirms this interpretation and suggests
that the Butcher-Oemler effect is a indicator of dynamical evolution as
well as color evolution.

\acknowledgments
The authors wish to thank the director and staff of the Lowell Observatory
for granting time for this project. Financial support from the Austrian
Fonds zur Foerderung der Wissenschaftlichen Forschung is gratefully
acknowledged.  We also wish to thank C. Leitherer for synthetic spectra of
starburst galaxies and R. Kennicut, A. De Bruyn and M. Ashby for numerical
values of measured spectra of galaxies.  This research has made use of the
NASA/IPAC Extragalactic Database (NED) which is operated by the Jet
Propulsion Laboratory, California Institute of Technology, under contract
with the National Aeronautics and Space Administration.

\figcaption[] {Two color diagrams for the Gunn-Stryker standard
stars.  Circles are main sequence stars, solid symbols are stars of
luminosity class IV and stars are stars of luminosity class III. The solid
line represents the position of metal poor stars, see text.  The reddening
vector for one magnitude of extinction is shown. \label{fig1}}

\figcaption[] {The $mz$ index diagrams for the Gunn-Stryker 
standard
stars.  Symbols are the same as Figure 1. \label{fig2}}

\figcaption[] {Two color diagrams for the synthetic colors of
galaxies with spectrophotometry from the literature.  The four classes of
galaxies; ellipticals, spirals, Seyferts and starbursts, are shown with
the reddest objects being ellipticals and the bluest being Seyferts in
continuum  emission.  The position of a B0I star and 3 magnitudes of
reddening are also shown.  Note that starburst galaxies following the
reddening curves from a young or intermediate age stellar population.
\label{fig3}}

\figcaption[] {The $mz$ index diagrams for the synthetic colors of
galaxies with spectrophotometry from the literature.  Starburst galaxies
occupy a specific region in the $mz$ plane with values at least less than
$-0.2$. \label{fig4}}

\figcaption[] {The starburst models of Leitherer and Heckman 
combined
with a typical elliptical population.  Three combinations are shown; equal
luminosity between elliptical and starburst maximum (solid line), one
magnitude fainter for the starburst (dotted line), one magnitude brighter
(dashed line).  The $mz$ index is the most sensitive to old starburst
detection.  The 4000{\AA} break color, $uz-vz$ is the least sensitive to
an aging starburst. \label{fig5}}

\figcaption[] {The starburst models of Leitherer and Heckman 
combined
with a typical spiral population.  Lines are the same as in Figure 5.  The
colors of the starburst approach normal spiral colors in only
$5\times10^7$ yrs. \label{fig6}}

\figcaption[] {Isophotal maps of the Arp/Markarian sample.  All
images are $yz$ band with aperture photometry regions from Table 2
marked. (This figure has been excluded in the preprint version to save
transfer time.) \label{fig7}}

\figcaption[] {Surface brightness profiles for Arp 169 A and B.  Note
the $r^{1/4}$ shape (straight line) typical for an elliptical galaxy.
\label{fig8}}

\figcaption[] {Two color diagrams for the Arp/Markarian sample.
Regions from Figure 3 are marked for three galaxy classes: ellipticals,
Seyferts and starbursts.  Different sized apertures are connected by
lines. \label{fig9}}

\figcaption[] {The $mz$ diagrams for the Arp/Markarian sample.  
The
dotted line is the equal luminosity starburst model from Figure 5.  Note
that a majority of the Arp/Markarian sample lies in the starburst region
of the diagram. \label{fig10}}

\figcaption[] {Histograms for the $mz$ values of ellipticals,
spirals, Seyferts and starburst galaxies are compared to the data from
intermediate and high redshift cluster galaxies.  The Butcher-Oemler
galaxies ($z>0.6$ blue galaxies) display a similar color distribution to
the present-day spiral sample with an additional blue tail due to a small 
fraction of starburst galaxies. \label{fig11}}

\clearpage
\begin{figure}
\epsscale{1.8}
\plotone{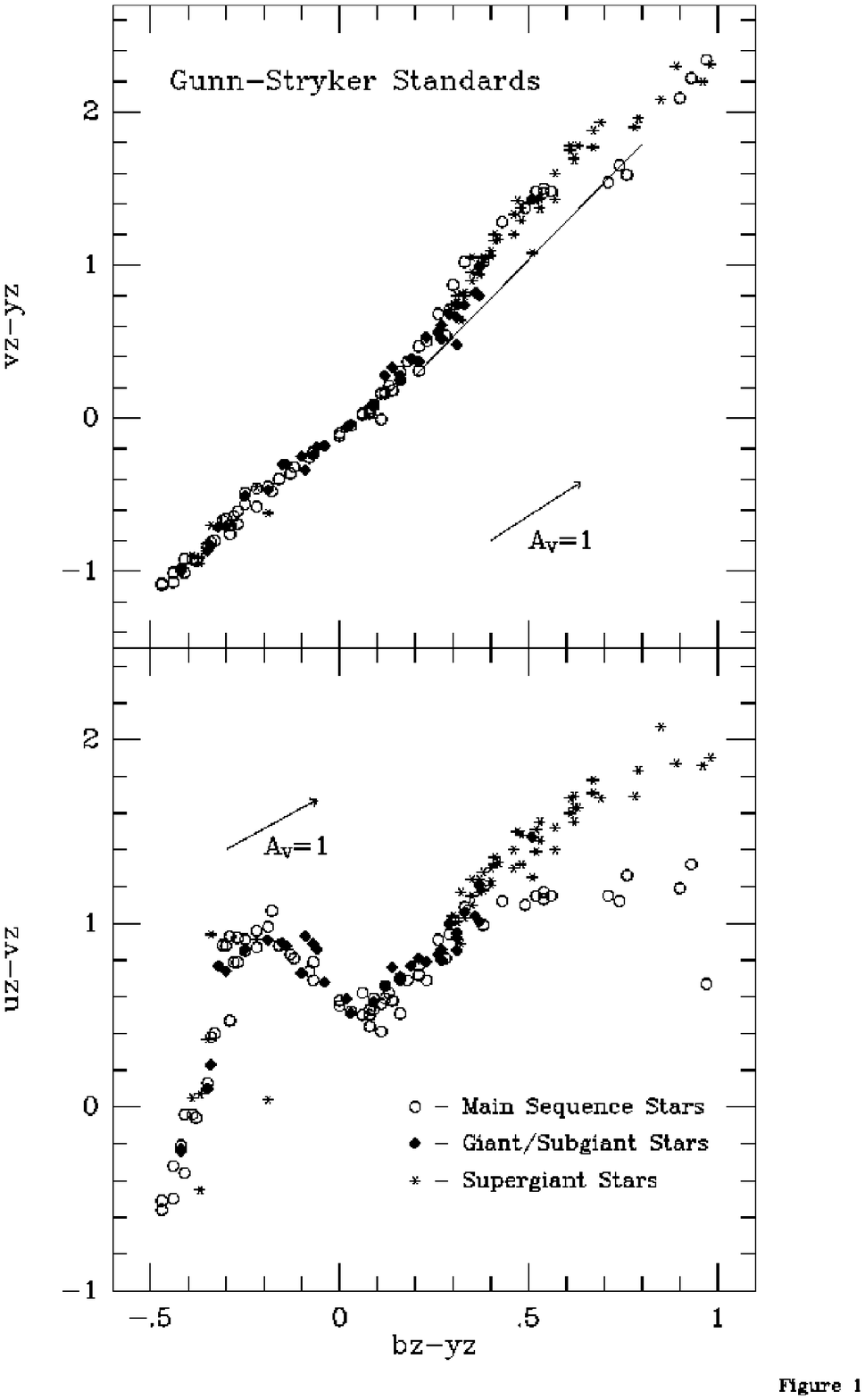}
\end{figure}
\clearpage
\begin{figure}
\epsscale{1.8}
\plotone{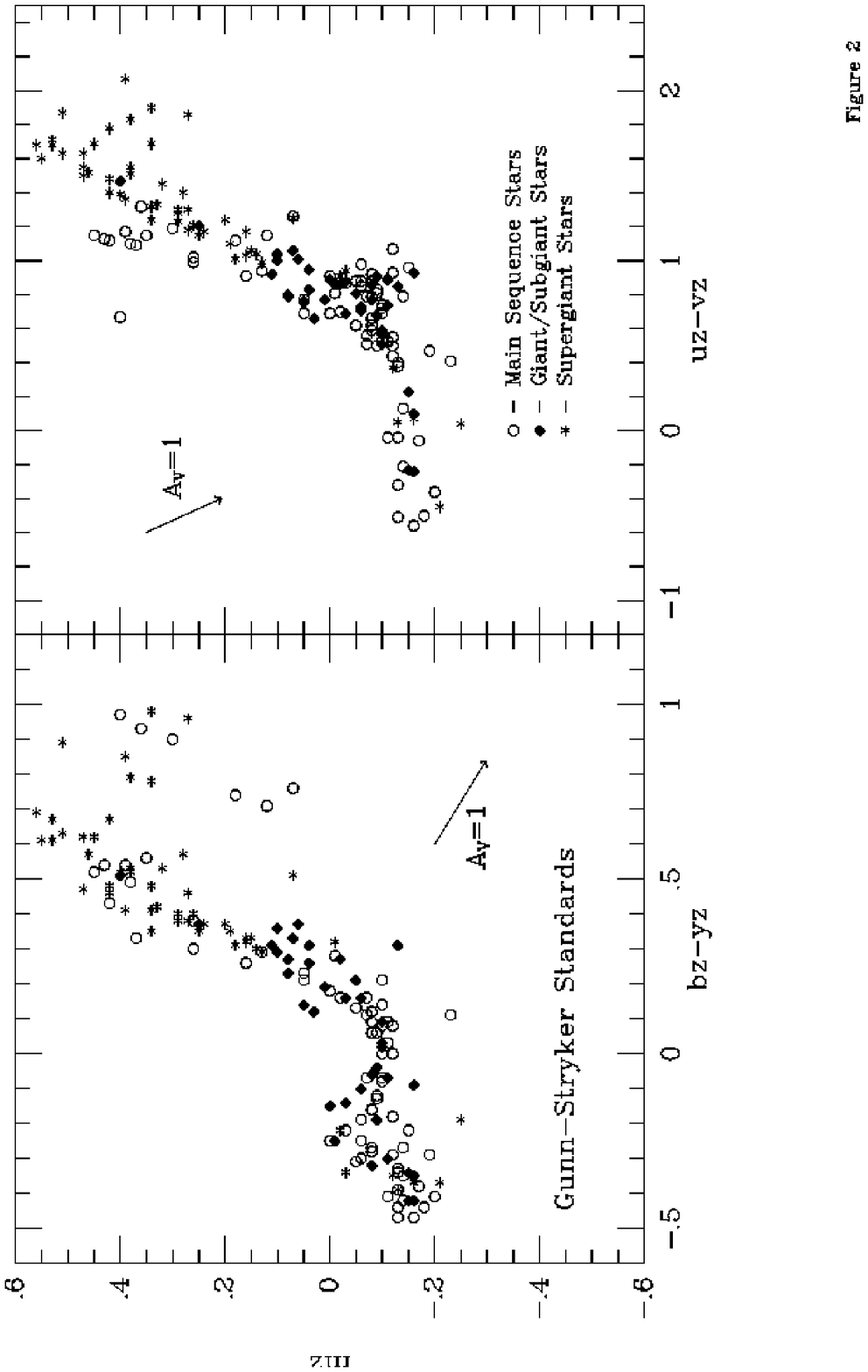}
\end{figure}
\clearpage
\begin{figure}
\epsscale{1.8}
\plotone{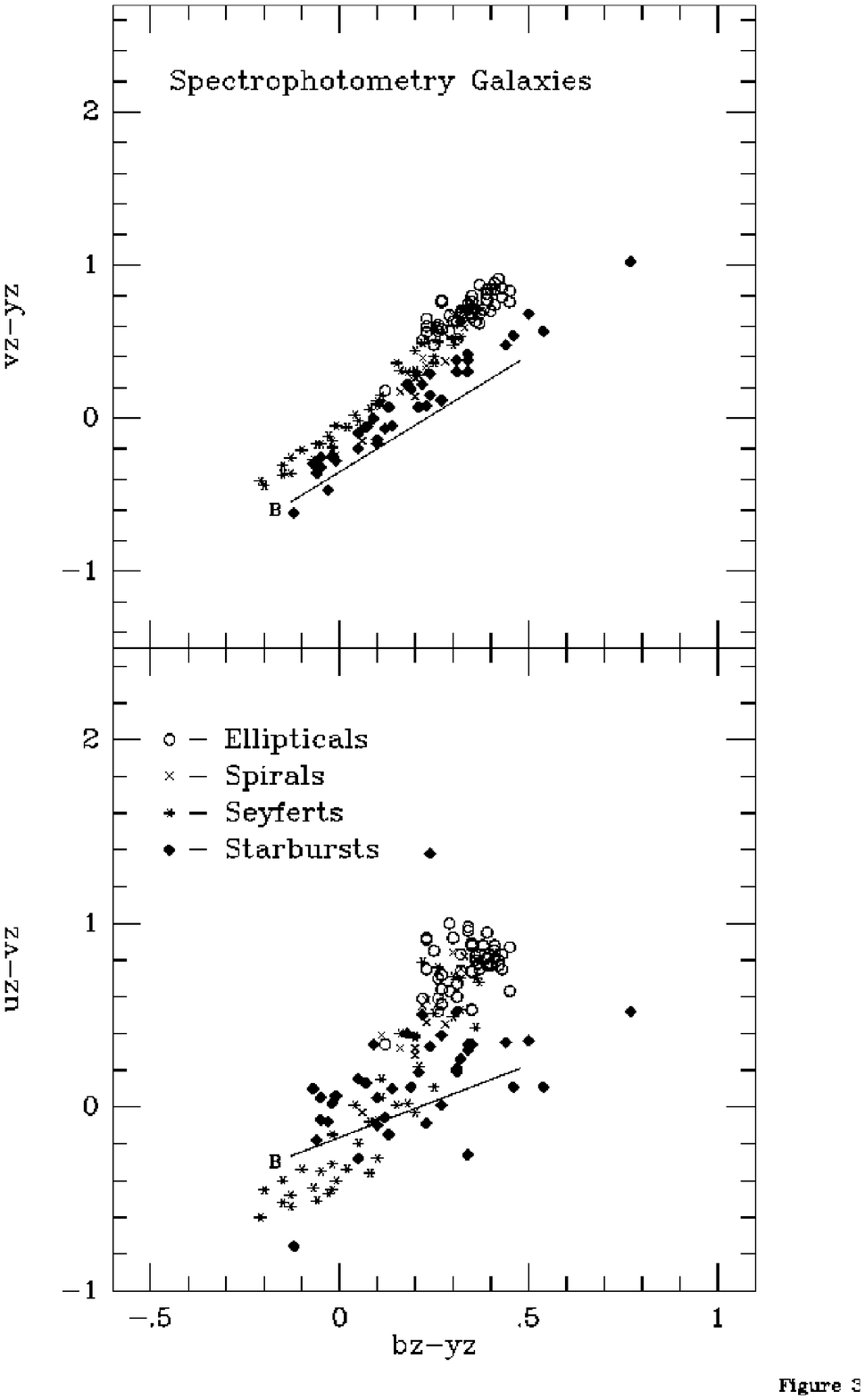}
\end{figure}
\clearpage
\begin{figure}
\epsscale{1.8}
\plotone{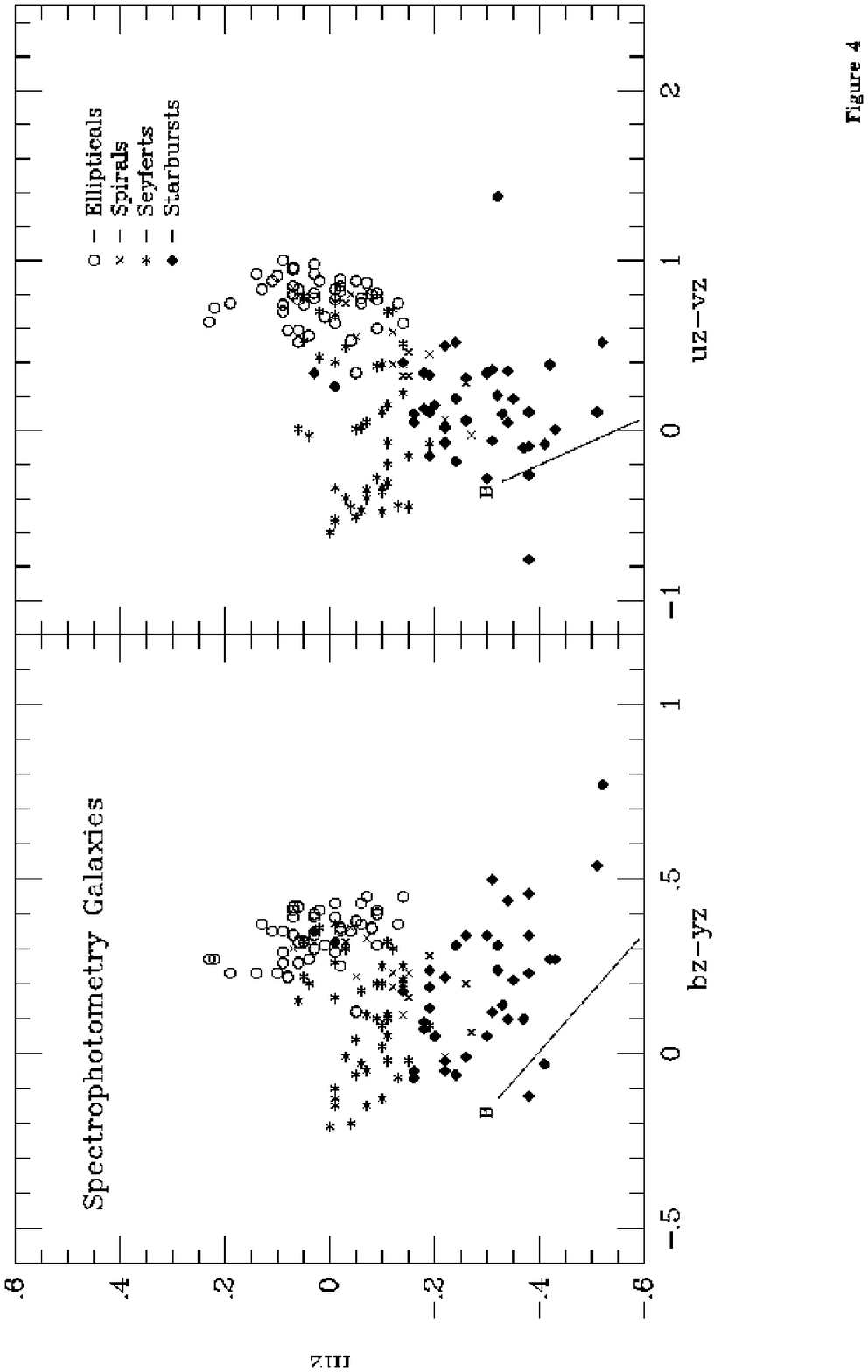}
\end{figure}
\clearpage
\begin{figure}
\epsscale{1.8}
\plotone{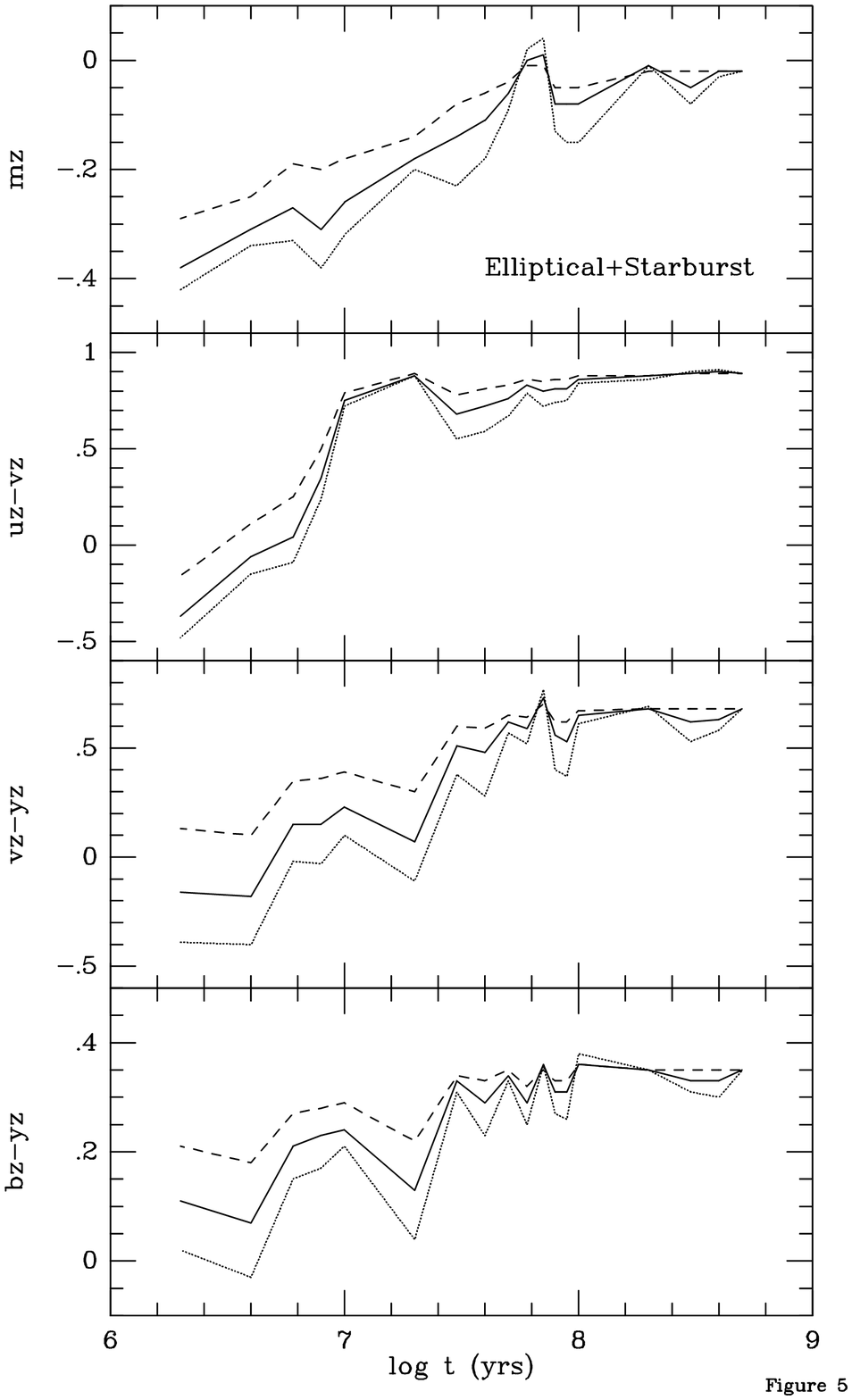}
\end{figure}
\clearpage
\begin{figure}
\epsscale{1.8}
\plotone{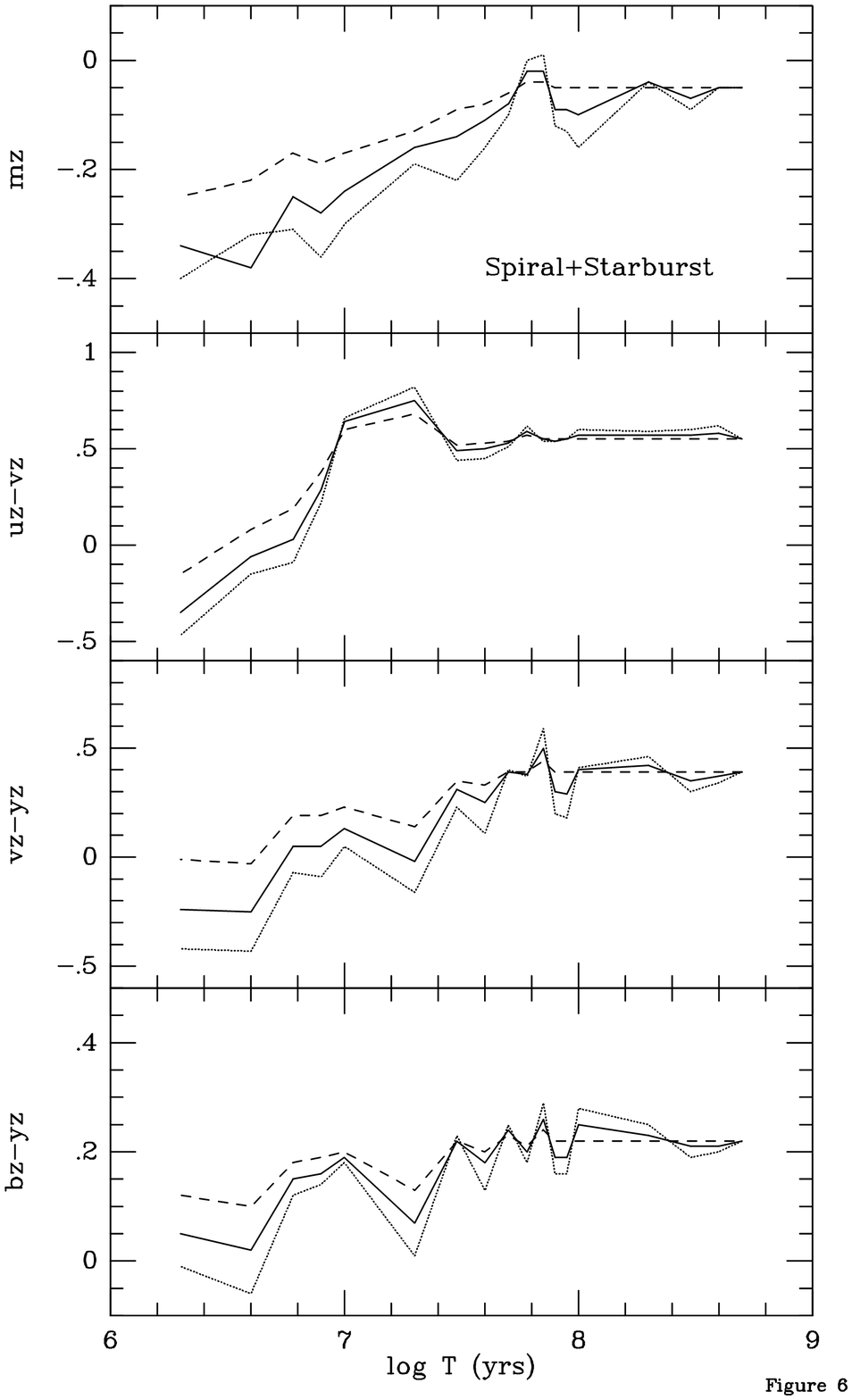}
\end{figure}
\clearpage
\begin{figure}
\epsscale{1.8}
\plotone{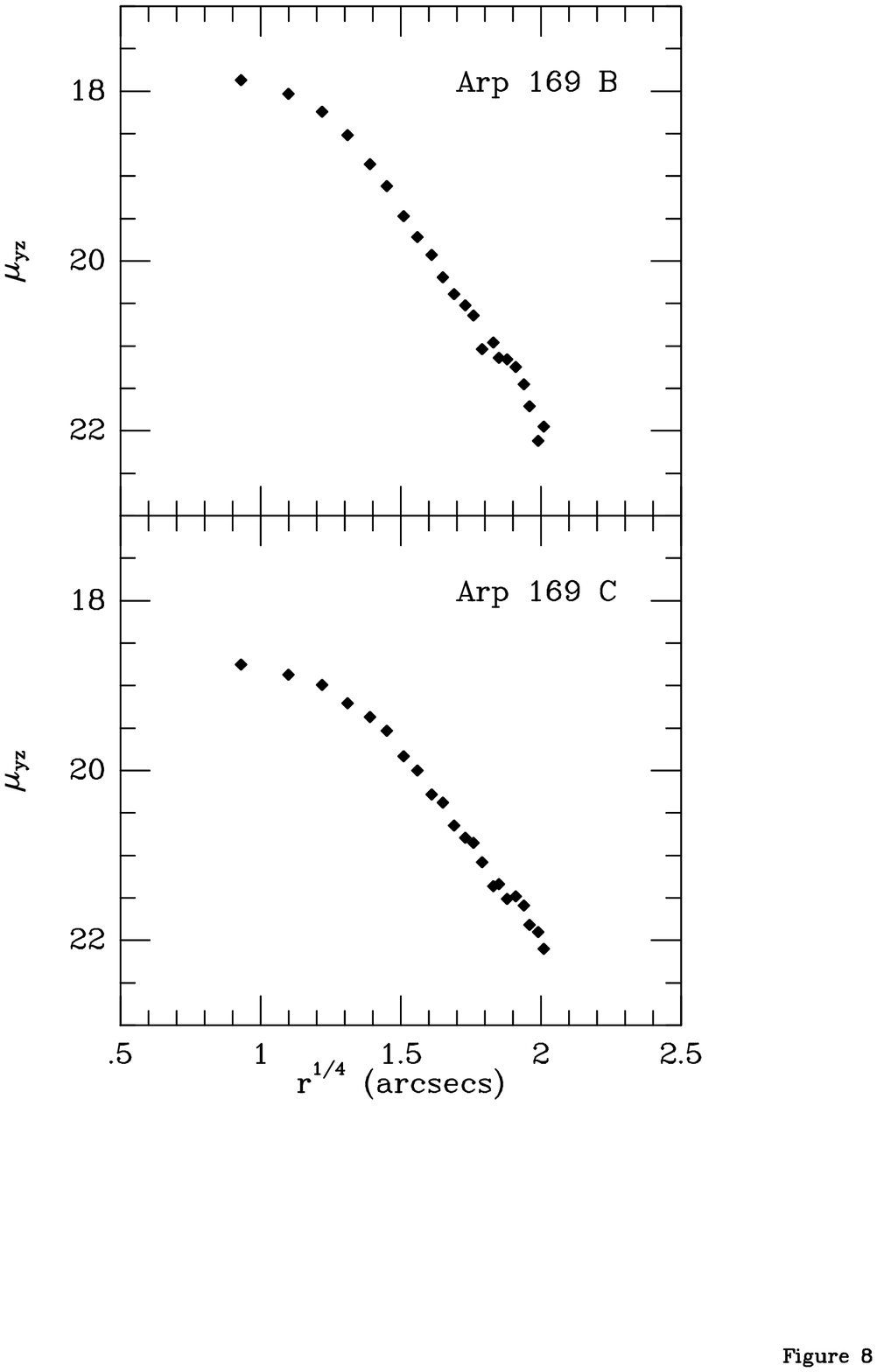}
\end{figure}
\clearpage
\begin{figure}
\epsscale{1.8}
\plotone{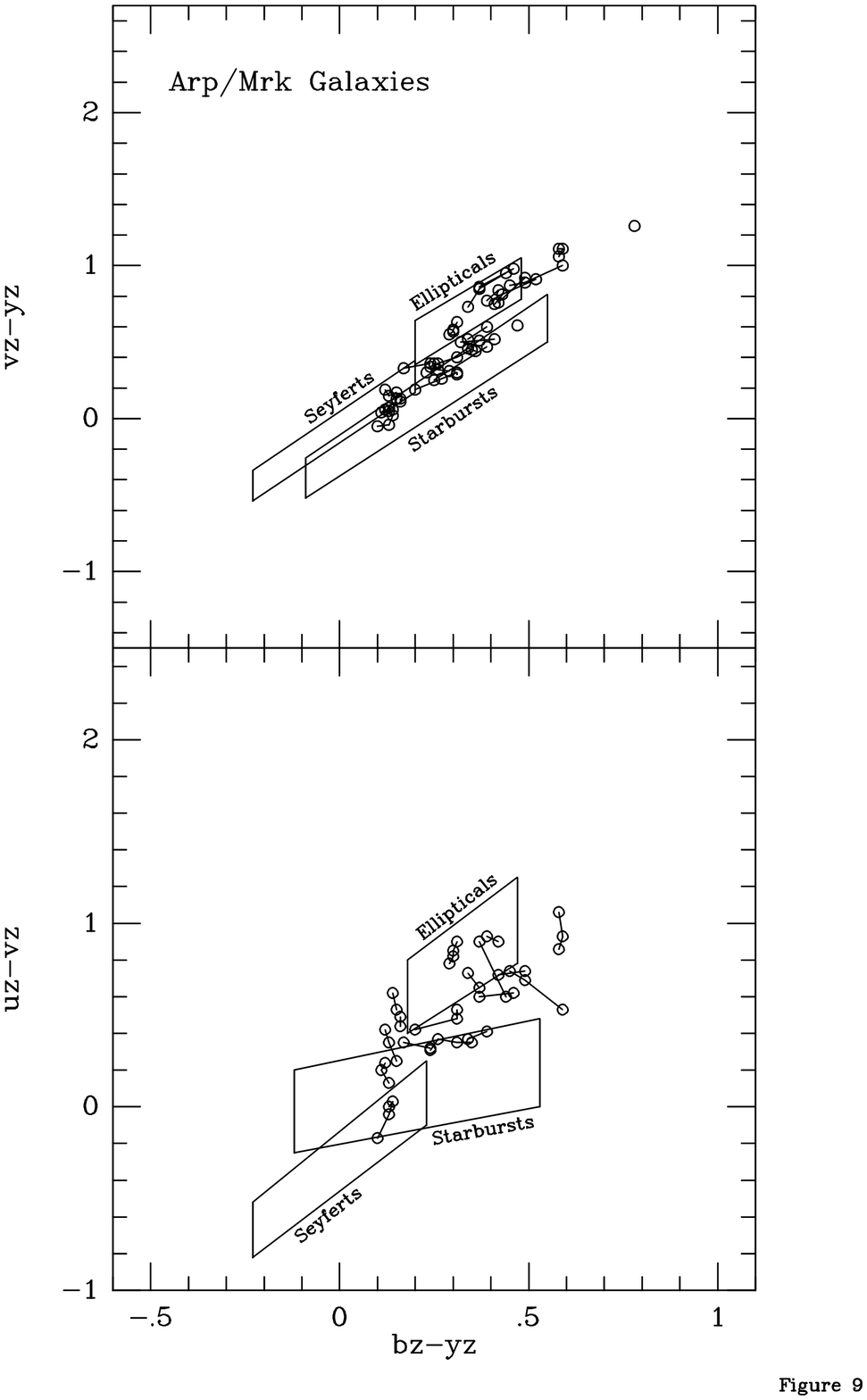}
\end{figure}
\clearpage
\begin{figure}
\epsscale{1.8}
\plotone{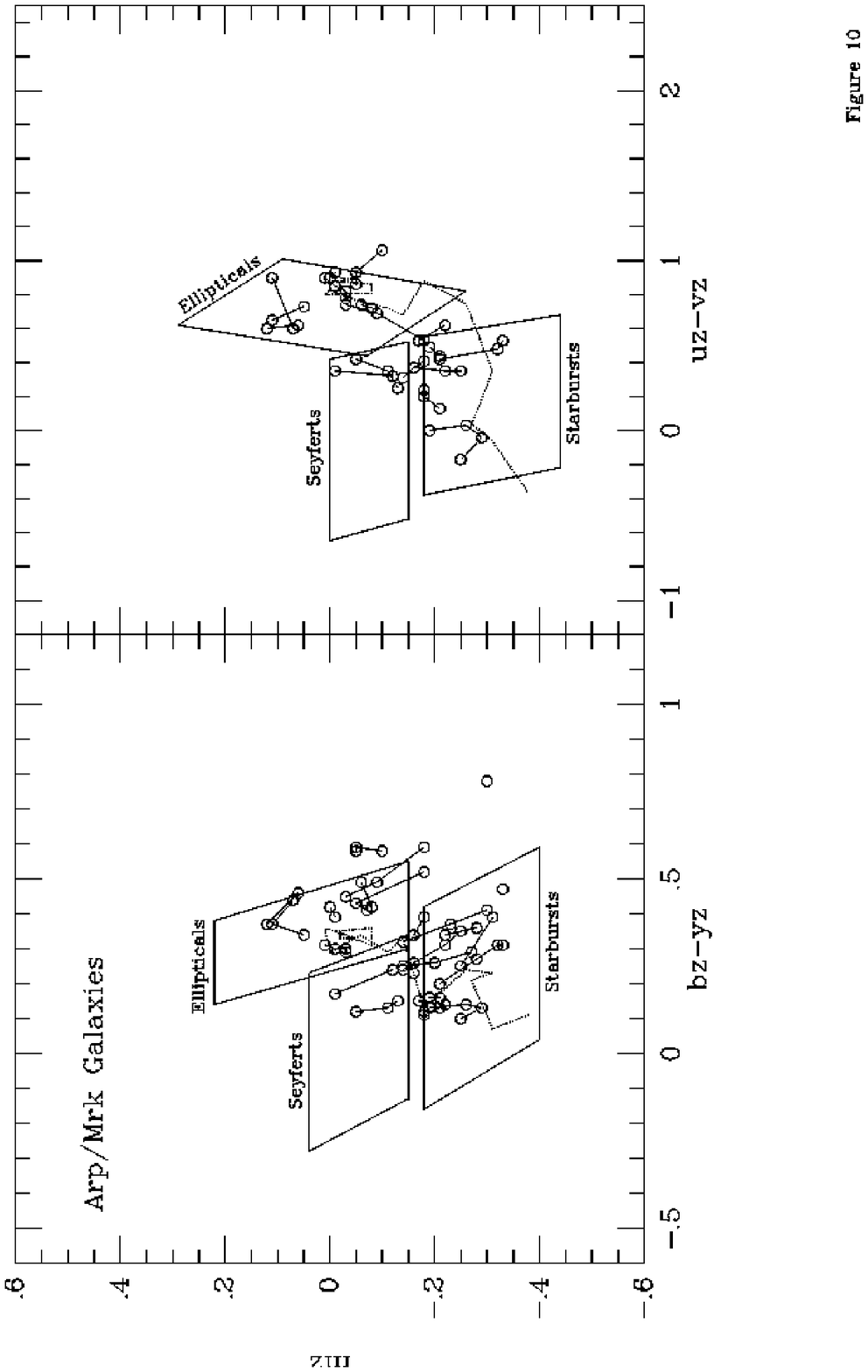}
\end{figure}
\clearpage
\begin{figure}
\epsscale{1.8}
\plotone{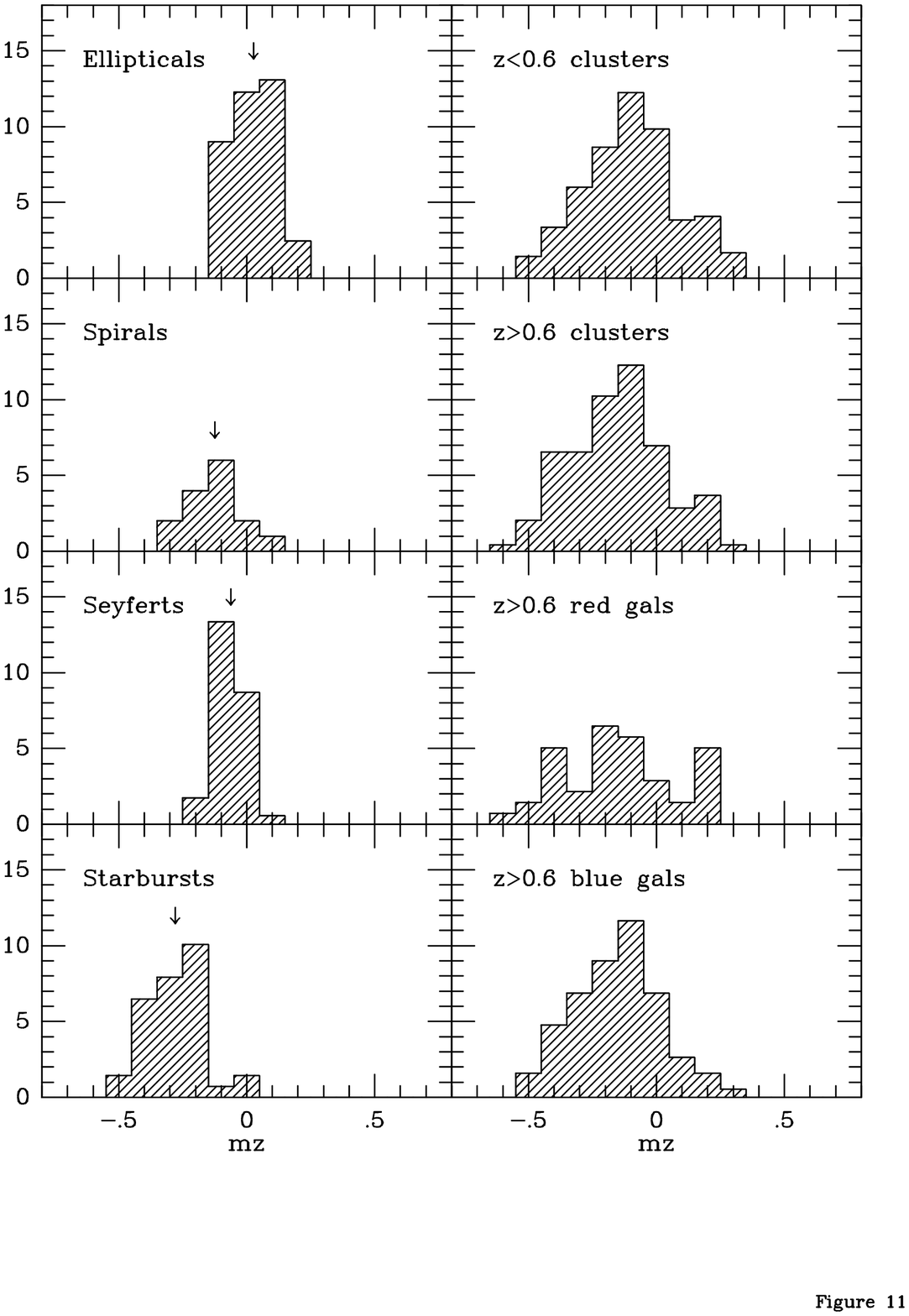}
\end{figure}
\end{document}